\documentclass[%
showpacs,preprintnumbers,
amsmath,amssymb,
 aps,
prl,
twocolumn,
]{revtex4-1}

\usepackage{graphicx}
\usepackage{dcolumn}
\usepackage{bm}
\usepackage{hyperref}

\begin{document}


\title{Thermoelectric detection of ferromagnetic resonance of a nanoscale ferromagnet}

\author{F. L. Bakker}
\email{f.l.bakker@rug.nl} 
\author{J. Flipse}
\author{A. Slachter}
\author{D. Wagenaar}
\author{B. J. van Wees}
\affiliation{Physics of Nanodevices, Zernike Institute for Advanced Materials, University
of Groningen, The Netherlands}%

\date{\today}

\begin{abstract}
We present thermoelectric measurements of the heat dissipated due to ferromagnetic resonance of a Permalloy strip. A microwave magnetic field, produced by an on-chip coplanar strip waveguide, is used to drive the magnetization precession. The generated heat is detected via Seebeck measurements on a thermocouple connected to the ferromagnet. The observed resonance peak shape is in agreement with the Landau-Lifshitz-Gilbert equation and is compared with thermoelectric finite element modeling. Unlike other methods, this technique is not restricted to electrically conductive media and is therefore also applicable to for instance ferromagnetic insulators.
\end{abstract}

\pacs{76.50.+g, 75.78.-n, 72.15.Jf, 85.80.Fi}

\maketitle
Thermal effects in ferromagnetic materials are subject to extensive research since the discovery of the spin-Seebeck effect \cite{Uchida2008, Jaworski2010, Uchida2010}. Recently, spin dynamics and (spin-) caloritronics, two popular branches of spintronics, started to come together as spin pumping induced by spin dynamics has been proposed as the origin of the spin-Seebeck effect \cite{Xiao2010,Adachi2011}. Magnetization dynamics has been studied thoroughly in magnetic systems as it is an important mechanism for future spintronic applications, e.g. for microwave generators \cite{Kiselev2003,Houssameddine2007} and spin sources via spin pumping \cite{Brataas2002, Tserkovnyak2002a}. However, dissipation mechanisms that accompany magnetization dynamics, and cause local heating, are still not fully understood \cite{Tserkovnyak2002,Brataas2008}.

Here, we focus on a new aspect, the coupling between magnetization dynamics and the generation of heat. We deduce from thermoelectric measurements on a Permalloy (Py) island  the heat dissipation during ferromagnetic resonance. This on-chip detection technique, based on the Seebeck effect, offers a novel method for characterizing ferromagnetic resonance and hence is distinctly different from other techniques, such as scanning thermal microscopy \cite{Meckenstock2008,Sakran2004}. Due to the thermal detection, electrical contact to the ferromagnet is in principle not required. Hence, this method allows for FMR measurements on non-conductive materials like ferromagnetic insulators.

When a ferromagnet is brought into resonance, energy is absorbed from the applied microwave field. This energy causes the magnetization $\vec{M}$ to precess around an effective field $\vec{H}$ and the motion is well described by the Landau-Lifshitz-Gilbert (LLG) equation $d\vec{M}/dt = -\gamma \vec{M} \times \vec{H} + (\alpha/M_s) \vec{M} \times d\vec{M}/dt$ with $\gamma$ = 176 GHz/T the gyromagnetic ratio. The last term in the LLG equation describes the damping of the magnetization towards the direction of the effective field $\vec{H}$, using the phenomenological damping parameter $\alpha$. This process is purely dissipative and converts magnetostatic energy into heat. During ferromagnetic resonance (FMR) this continuous dissipation leads to heating of the ferromagnetic material.

\begin{figure}[b]
\includegraphics[width=6.5cm]{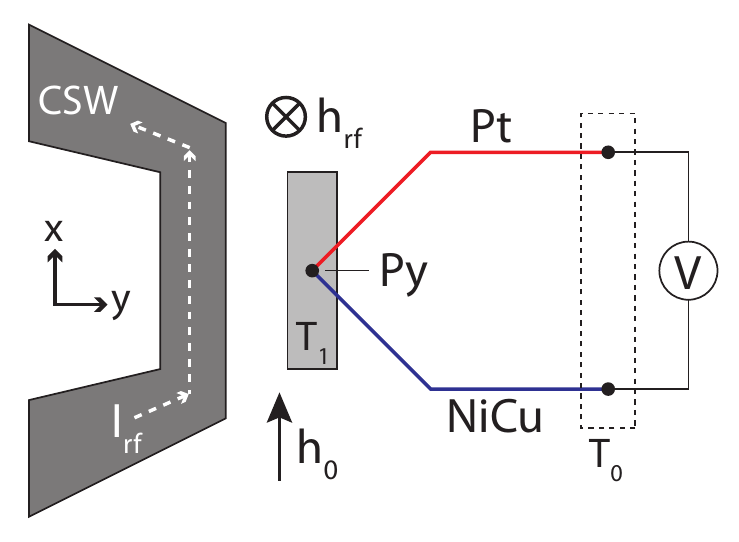}
\caption{\label{fig:concept} (color online) Concept of the thermoelectric detection of ferromagnetic resonance. A coplanar strip waveguide generates a microwave magnetic field $h_{rf}$ in the $\hat{z}$ direction acting on a small ferromagnetic strip. A static magnetic field $h_0$ is applied along the $\hat{x}$ axis. When the strip is brought into resonance, it absorbs energy from the field which is dissipated as heat. The dissipation is detected by a Pt-NiCu thermocouple which probes the temperature $T_1$ of the ferromagnet with respect to a reference temperature $T_0$ via the Seebeck effect.}
\end{figure}

In this experiment, we measured the temperature of a ferromagnet while subject to a microwave magnetic field. A ferromagnetic strip is placed close to the shortened end of a coplanar strip waveguide (CSW) as shown in Fig.~\ref{fig:concept}. Microwave power is applied to the CSW, leading to an out of plane rf magnetic field. A static magnetic field $h_0$ is applied along the easy axis of the magnet. In addition, a thermocouple consisting of a NiCu and Pt wire is connected to the ferromagnet by a Au bridge. In this way, the temperature can be measured by making use of the Seebeck effect. The Seebeck effect describes the generation of a voltage due to a temperature gradient, $\nabla V = - S \nabla T$, with $S$ the material dependent Seebeck coefficient. The voltage that develops across the Pt wire is different than the voltage that develops across the NiCu wire, leading to a nonzero voltage between the two wires. This thermovoltage scales with the Seebeck coefficients $(S_{\text{Pt}} - S_{\text{NiCu}})$ and the temperature difference $(T_1 - T_0)$. Note that NiCu is chosen because of its relatively high Seebeck coefficient ($S_{\text{NiCu}} = \text{-40} \mu V/K$ and $S_{\text{Pt}} = \text{-5} \mu V/K$). In the following, we calculate the temperature rise during FMR from the dissipated power.

The energy of a ferromagnetic particle in a magnetic field, the Zeeman energy, is given by:
\begin{equation}
E  = - \int_V \vec{M} \cdot \vec{B} dV
\end{equation}
with $\vec{M}$ the magnetization, $\vec{B} = \mu_0 \left( \vec{H}_{\text{ext}} + \vec{H}_{D}/2 \right)$ the sum of the externally applied magnetic field and the demagnetizing field and $V$ the volume of the particle \cite{coey2010magnetism}. Here, $\vec{H}_{D}$ is divided by two to compensate for counting each volume twice in the integral. For this experiment, a static magnetic field, $h_0$, is applied in the $\hat{x}$ direction and a driving rf field, $h_{rf} \cos \omega t$, in the $\hat{z}$ direction, making $\vec{B}  = \left( h_0 - N_xm_x/2, -N_ym_y/2, h_{rf} \cos \omega t -N_zm_z/2 \right)$ with $N_x$, $N_y$ and $N_z$ the demagnetization factors. The dissipation energy can now be calculated from the time derivative of $E$, assuming a uniform $\vec{M}$ and $\vec{B}$:
\begin{equation}\label{eq:dEdt}
\frac{dE}{dt}  = - V \left( \frac{d\vec{M}}{dt} \cdot \vec{B} + \vec{M} \cdot \frac{d\vec{B}}{dt} \right)
\end{equation}
where the first part of Eq. \ref{eq:dEdt} expresses the dissipation due to the magnetization motion and the second part the energy absorbed from the microwave field. In equilibrium, the absorption of energy from the microwave field equals the dissipation, $\left\langle dE/dt \right\rangle = 0 $, leading to heating of the ferromagnet. In order to find an expression for the dissipated power, we use a procedure similar to Ref. \cite{Costache2006a} where the magnetization dynamics is described by the linearized LLG equation. We assume that for small angle precessional motion $dm_x/dt = 0$ such that $m_x$ is constant and the solution to the Landau-Lifshitz-Gilbert (LLG) equation can be written in terms of the sum of in-phase and out-of-phase susceptibilities. The components $m_y$ and $m_z$ are now defined as $m_y = \chi_y' \omega_1 \cos \omega t + \chi_y'' \omega_1 \sin \omega t$ and $m_z = \chi_z' \omega_1 \cos \omega t + \chi_z'' \omega_1 \sin \omega t$ with:
\begin{equation}\label{eq:susc}
\begin{split}
\chi_y' = \frac{\alpha \omega^2 ( \omega_{y} + \omega_{z} )}{(\omega^2-\omega_y \omega_z)^2 + (\alpha\omega)^2(\omega_y+\omega_z)^2} \\
\chi_y'' = \frac{\omega (\omega^2 - \omega_y\omega_z)}{(\omega^2-\omega_y \omega_z)^2 + (\alpha\omega)^2(\omega_y+\omega_z)^2}
\end{split}
\end{equation}
Here, $\omega$ is the frequency of the driving rf field, $\omega_y = \gamma(h_0 - (N_x - N_y)m_s)$, $\omega_z = \gamma(h_0 - (N_x - N_z)m_s)$, $\omega_1 = \gamma h_{rf}$ and $m_s$ the saturation magnetization. The $\chi_z'$ and $\chi_z''$ are related via $\chi_z' = \alpha \chi_y' - \omega_y/\omega \chi_y''$ and $\chi_z'' = \alpha \chi_y'' + \omega_y/\omega \chi_y'$, respectively. With these expressions for $m_x$, $m_y$, $m_z$ and $B_x$, $B_y$, $B_z$ one can easily find its time derivatives and calculate the relevant dot product of Eq. \ref{eq:dEdt}. We are not interested in high frequency variations in the dissipated power and hence, average out all contributions that vary with a frequency $\omega$ or 2$\omega$ and find:
\begin{equation}\label{eq:disspower}
\left\langle \frac{dE}{dt} \right\rangle_{\text{dissipation}} = \frac{\chi_z'' \omega_1^2 \omega V}{2 \gamma}
\end{equation}
From this expression we can deduce that the resonance peak shape of the dissipated power is determined by $\chi_z''$ and scales with the applied microwave field and frequency. In order to convert this power into a temperature rise, we make use of 3D finite element thermoelectric modeling. For details about the modeling we refer to earlier publications \cite{Bakker2010, Slachter2011}.

\begin{figure}[t]
\includegraphics[width=8.8cm]{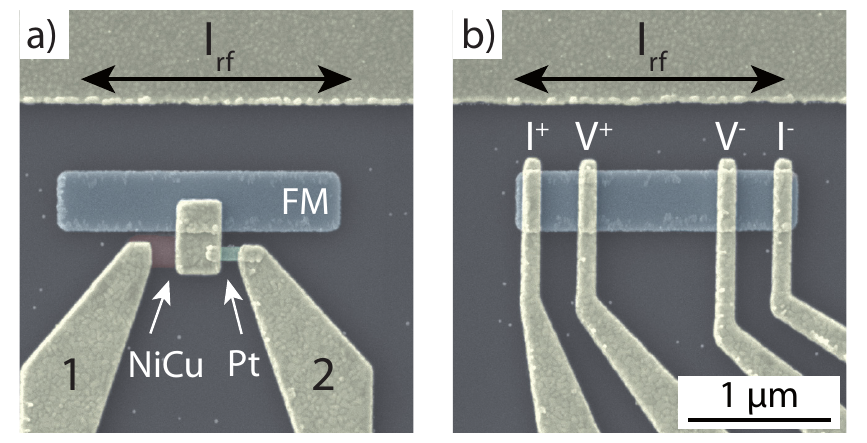}
\caption{\label{fig:sem} (color online) a) Scanning electron microscope (SEM) image of the thermoelectric FMR device. b) Image of a device with four contacts for dc AMR detection of FMR.}
\end{figure}

The samples are fabricated using a three-step electron beam lithography process on top of a thermally oxidized Si substrate. A scanning electron microscope (SEM) image of the investigated devices is shown in Fig. \ref{fig:sem}. The devices consist of a 50nm thick Py strip (2 $\mu$m $\times$ 400~nm) close to a 100~nm thick Au coplanar strip waveguide. The CSW is made using an optical lithography process. In the thermoelectric device (Fig.~\ref{fig:sem}a), there are two 40~nm thick contacts (Pt and NiCu) forming a thermocouple. The Py island is connected to the thermocouple by a highly thermal conductive Au contact (thickness: 120~nm). The other side of the thermocouple is connected by 120~nm thick Au contacts to the bonding pads. In the case of the anisotropic magnetoresistance (AMR) device (Fig.~\ref{fig:sem}b), four 120~nm thick Au contacts directly connect to the Py strip. The NiCu is deposited by DC sputtering to preserve the original alloy composition (45\% Ni, 55\% Cu). To avoid lift-off problems a double-layer resist technique with a large undercut (PMMA-MA and PMMA 950K) is used. The Au, Pt and Py are deposited using an e-beam evaporator (base pressure 1 $\times$ 10$^{-7}$~mbar) and a single layer resist (PMMA 950K). Prior to the Au deposition, the NiCu, Pt and Py surfaces are cleaned with Ar ion milling.

For the measurements, we have used a frequency modulation method to obtain a better signal to noise ratio and to remove background voltages due to heating of the CSW short. The microwave field frequency is alternated between two different values with a separation of 5 GHz. A lock-in amplifier, tuned to the same frequency (17~Hz), measures the difference in dc voltage across contacts 1 and 2 (Fig.~\ref{fig:sem}a) between the two frequencies ($V = V_{\text{f=high}} - V_{\text{f=low}}$). Because of the large separation between $V_{\text{f=low}}$ and $V_{\text{f=high}}$, they can not both fulfill the resonance condition at a specific magnetic field. With this method, one effectively measures the difference in the Seebeck or AMR voltage when the ferromagnet is in- and off-resonance. All measurements were performed at room temperature.

\begin{figure}[t]
\includegraphics[width=8.5cm]{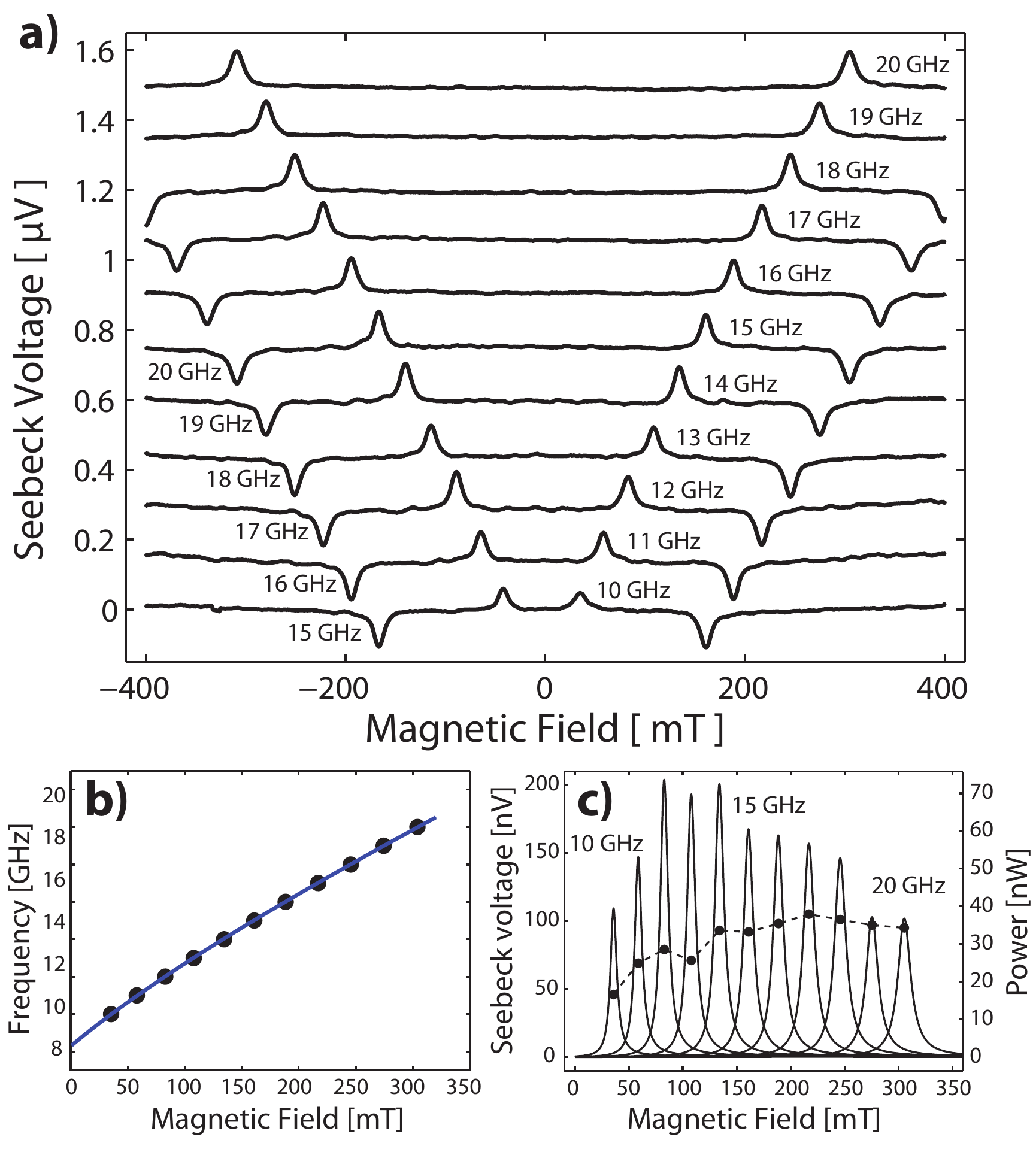}
\caption{\label{fig:data1} (color online) a) Series of Seebeck voltage versus magnetic field measurements for 11 different frequencies. The traces are offset by 150~nV for clarity. Due to the modulation technique using two driving frequencies that are 5 GHz apart, peaks and dips are observed at the resonance fields for both frequencies. b) Frequency versus the magnetic field at the center of the resonance peak. The line corresponds to a fit of the Kittel equation. c) Generated power and corresponding Seebeck voltage calculated using Eq.~\ref{eq:disspower} and thermoelectric finite-element modeling for multiple frequencies. The measured peak heights of a) are indicated by the black dots.}
\end{figure}

Fig. \ref{fig:data1}a shows the measured Seebeck voltage as a function of magnetic field for different rf field frequencies (10 - 20 GHz) for 12 dBm rf power. The position of the peaks and dips correspond to the resonance field for $f_{\text{low}}$ and $f_{\text{high}}$, respectively. We have plotted the peak position as a function of the applied rf frequency in Fig. \ref{fig:data1}b and found peak heights ranging from 46~nV at 10~GHz to 105~nV at 17~GHz. For a uniform precessional mode, the resonance field is related to $\omega$ by the Kittel equation \cite{Kittel1995}:
\begin{equation}\label{eq:Kittel}
\omega^2 = \gamma^2(h_0-(N_x - N_y)m_s)(h_0 - (N_x - N_z)m_s)
\end{equation}
The line corresponds to a fit of Eq. \ref{eq:Kittel} and the almost perfect fit confirms the uniform precessional motion. We obtained the following fitting parameters: $N_x = 0.01$, $N_y = 0.09$, $N_z = 0.90$ and $\mu_0 m_s = 1.11$~T. These parameters have been used to calculate the dissipated power of Eq.~\ref{eq:disspower} for different frequencies. However, in order to do this accurately one first need to determine the magnitude of the rf magnetic field experimentally. 

\begin{figure}[ht]
\includegraphics[width=8.5cm]{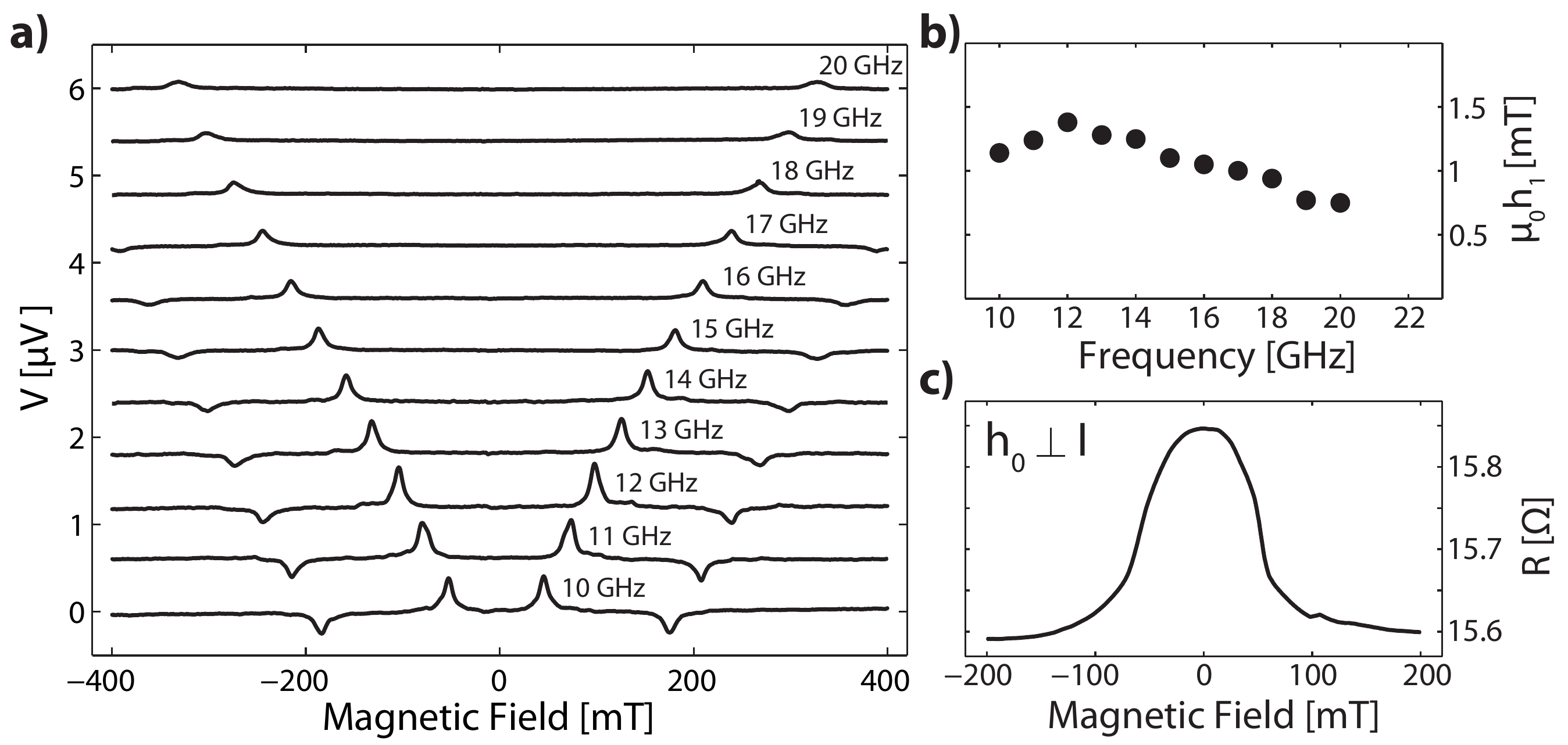}
\caption{\label{fig:data2} (color online) a) AMR voltage vs. magnetic field. Peaks and dips are observed for the resonance fields of the two driving frequencies (5 GHz apart). The different traces are offset for clarity reasons. b) The magnitude of the rf field is extracted from the peak height of the resonance ($V = I_{\text{dc}} \Delta R \sin^2 \theta_c$). c) Anisotropic magnetoresistance measurement with $\vec{I}$ and $\vec{B}$ perpendicularly aligned. }
\end{figure}

To obtain the correct experimental value for the magnitude of the rf magnetic field, we have measured the dc anisotropic magnetoresistace (AMR) in a dedicated device with a four-terminal geometry (shown in Fig.~\ref{fig:sem}b). The AMR effect describes the dependence of the resistance on the angle $\theta$ between the current $\vec{I}$ and the direction of magnetization $\vec{M}$ by $R = R_0 - \Delta R \sin^2 \theta$, where $R_0$ is the resistance of the strip when $\vec{I}$ and $\vec{M}$ are parallel, and $\Delta R$ the difference in resistance between the parallel and perpendicular alignment of $\vec{I}$ and $\vec{M}$. A measurement of $R$ as a function of a perpendicular applied magnetic field is plotted in Fig.~\ref{fig:data2}c. From this measurement, we determined the magnitude of the AMR effect and found $\Delta R / R_0 = 1.5~\%$. For a steady resonant precession, the average cone angle $\theta_c$ of the precession can now be extracted from the observed AMR voltage.

Fig.~\ref{fig:data2}a displays the AMR voltage versus magnetic field for different microwave field frequencies. For this measurement, we have used a dc current $I_{\text{dc}}$ of 300~$\mu$A. The obtained voltage now corresponds to the dc current multiplied with the resistance change, being $V = I_{\text{dc}} \Delta R \sin^2 \theta_c$, and the precession angle can be extracted. Using Eq.~\ref{eq:susc} and an expression for the average cone angle:
\begin{equation}
\left\langle \theta^2_c \right\rangle  = \omega_1^2(\chi_y'^{2} + \chi_y''^{2} + \chi_z'^{2} + \chi_z''^{2})/2
\end{equation}
one can deduce $h_{rf} = \omega_1/\gamma$ from a fit of the measured peak height, and the result is plotted in Fig.~\ref{fig:data2}b for a microwave power of 12 dBm. The field strength is found to decrease twofold when the frequency is increased from 10 to 20~GHz. We attribute this to frequency dependent attenuation of the microwave signal, leading to smaller rf fields at higher frequencies.

Now we can calculate, using finite element modeling in Comsol Multiphysics, the Seebeck voltage that is generated due to the heating of the ferromagnet. In this model we impose the constant heat flux, given by Eq.~\ref{eq:disspower}, through the top layer of the ferromagnet and solve the thermoelectric model \cite{Bakker2010,Slachter2011}. Both, the heat flux and the calculated Seebeck voltage are plotted in Fig.~\ref{fig:data1}c (solid lines) for multiple frequencies. Peak heights ranging from 98 till 197~nV are calculated. For comparison, the observed peak height of Fig. \ref{fig:data1}a is replotted in Fig.~\ref{fig:data1}c as black dots. 

For a fixed rf field strength, the Seebeck voltage should increase monotonically with frequency due to an increasing dissipation (see Eq.~\ref{eq:disspower}). However, because of the experimental variation in rf field strength for different frequencies, a specific relation between the calculated Seebeck voltage and the frequency is found (Fig.~\ref{fig:data1}c). The experimental data is in agreement with the calculations within a factor of two and follows partially the same trend. This discrepancy is attributed to small sample to sample variations in the rf field strength. Since the AMR measurements are performed on a separate device, the fields can differ for the thermoelectric device. Moreover, small shifts in the contact area of the thermocouple can lead to changes in the heat transport and hence, different thermo voltages.

Furthermore, circulating rf currents combined with an oscillating magnetoresistance at the same frequency can cause dc voltages via a rectifying effect and mimic the observed thermal behavior in our devices \cite{Costache2006a}. We have excluded these effects by using a similar device with an Au-Au thermocouple such that the Seebeck effect vanishes. For this device we observed a flat background voltage without peaks and dips. We note that thermal voltages can be of importance in other device geometries where ferromagnets are electrically connected to nonmagnetic metals. For example, detection of interface voltages that arise due to spin-pumping \cite{Wang2006,Costache2006b} cannot be easily distinguished from generated Seebeck voltages. Thorough temperature or material dependent measurements might offer a solution to discriminate between both mechanisms.

In conclusion, we have demonstrated a new thermoelectric detection technique for ferromagnetic resonance.  The observed resonance peaks are in good agreement with the LLG equation and thermoelectric finite element modeling. Additionally, this technique can be applied on the nanoscale and is not limited to conductive ferromagnetic media. Thermal detection offers a valid, alternative method for studying the dissipation, i.e. the Gilbert damping term, in nanoscale ferromagnetic islands. We hope that these results stimulate research for new physical effects that arises from coupling between magnetization dynamics and caloritronics.

We would like to acknowledge B. Wolfs, M. de Roosz and J.G. Holstein for technical assistance. This work is part of the research program of the Foundation for Fundamental Research on Matter (FOM) and supported by NanoLab and the Zernike Institute for Advanced Materials.


\begin{thebibliography}{20}%
\makeatletter
\providecommand \@ifxundefined [1]{%
 \@ifx{#1\undefined}
}%
\providecommand \@ifnum [1]{%
 \ifnum #1\expandafter \@firstoftwo
 \else \expandafter \@secondoftwo
 \fi
}%
\providecommand \@ifx [1]{%
 \ifx #1\expandafter \@firstoftwo
 \else \expandafter \@secondoftwo
 \fi
}%
\providecommand \natexlab [1]{#1}%
\providecommand \enquote  [1]{``#1''}%
\providecommand \bibnamefont  [1]{#1}%
\providecommand \bibfnamefont [1]{#1}%
\providecommand \citenamefont [1]{#1}%
\providecommand \href@noop [0]{\@secondoftwo}%
\providecommand \href [0]{\begingroup \@sanitize@url \@href}%
\providecommand \@href[1]{\@@startlink{#1}\@@href}%
\providecommand \@@href[1]{\endgroup#1\@@endlink}%
\providecommand \@sanitize@url [0]{\catcode `\\12\catcode `\$12\catcode
  `\&12\catcode `\#12\catcode `\^12\catcode `\_12\catcode `\%12\relax}%
\providecommand \@@startlink[1]{}%
\providecommand \@@endlink[0]{}%
\providecommand \url  [0]{\begingroup\@sanitize@url \@url }%
\providecommand \@url [1]{\endgroup\@href {#1}{\urlprefix }}%
\providecommand \urlprefix  [0]{URL }%
\providecommand \Eprint [0]{\href }%
\providecommand \doibase [0]{http://dx.doi.org/}%
\providecommand \selectlanguage [0]{\@gobble}%
\providecommand \bibinfo  [0]{\@secondoftwo}%
\providecommand \bibfield  [0]{\@secondoftwo}%
\providecommand \translation [1]{[#1]}%
\providecommand \BibitemOpen [0]{}%
\providecommand \bibitemStop [0]{}%
\providecommand \bibitemNoStop [0]{.\EOS\space}%
\providecommand \EOS [0]{\spacefactor3000\relax}%
\providecommand \BibitemShut  [1]{\csname bibitem#1\endcsname}%
\let\auto@bib@innerbib\@empty
\bibitem [{\citenamefont {Uchida}\ \emph {et~al.}(2008)\citenamefont {Uchida},
  \citenamefont {Takahashi}, \citenamefont {Harii}, \citenamefont {Ieda},
  \citenamefont {Koshibae}, \citenamefont {Ando}, \citenamefont {Maekawa},\
  and\ \citenamefont {Saitoh}}]{Uchida2008}%
  \BibitemOpen
  \bibfield  {author} {\bibinfo {author} {\bibfnamefont {K.}~\bibnamefont
  {Uchida}}, \bibinfo {author} {\bibfnamefont {S.}~\bibnamefont {Takahashi}},
  \bibinfo {author} {\bibfnamefont {K.}~\bibnamefont {Harii}}, \bibinfo
  {author} {\bibfnamefont {J.}~\bibnamefont {Ieda}}, \bibinfo {author}
  {\bibfnamefont {W.}~\bibnamefont {Koshibae}}, \bibinfo {author}
  {\bibfnamefont {K.}~\bibnamefont {Ando}}, \bibinfo {author} {\bibfnamefont
  {S.}~\bibnamefont {Maekawa}}, \ and\ \bibinfo {author} {\bibfnamefont
  {E.}~\bibnamefont {Saitoh}},\ }\href {\doibase 10.1038/nature07321}
  {\bibfield  {journal} {\bibinfo  {journal} {Nature}\ }\textbf {\bibinfo
  {volume} {455}},\ \bibinfo {pages} {778} (\bibinfo {year}
  {2008})}\BibitemShut {NoStop}%
\bibitem [{\citenamefont {Jaworski}\ \emph {et~al.}(2010)\citenamefont
  {Jaworski}, \citenamefont {Yang}, \citenamefont {Mack}, \citenamefont
  {Awschalom}, \citenamefont {Heremans},\ and\ \citenamefont
  {Myers}}]{Jaworski2010}%
  \BibitemOpen
  \bibfield  {author} {\bibinfo {author} {\bibfnamefont {C.~M.}\ \bibnamefont
  {Jaworski}}, \bibinfo {author} {\bibfnamefont {J.}~\bibnamefont {Yang}},
  \bibinfo {author} {\bibfnamefont {S.}~\bibnamefont {Mack}}, \bibinfo {author}
  {\bibfnamefont {D.~D.}\ \bibnamefont {Awschalom}}, \bibinfo {author}
  {\bibfnamefont {J.~P.}\ \bibnamefont {Heremans}}, \ and\ \bibinfo {author}
  {\bibfnamefont {R.~C.}\ \bibnamefont {Myers}},\ }\href {\doibase
  10.1038/nmat2860} {\bibfield  {journal} {\bibinfo  {journal} {Nature
  Materials}\ }\textbf {\bibinfo {volume} {9}},\ \bibinfo {pages} {898} (\bibinfo
  {year} {2010})}\BibitemShut {NoStop}%
\bibitem [{\citenamefont {Uchida}\ \emph {et~al.}(2010)\citenamefont {Uchida},
  \citenamefont {Xiao}, \citenamefont {Adachi}, \citenamefont {Ohe},
  \citenamefont {Takahashi}, \citenamefont {Ieda}, \citenamefont {Ota},
  \citenamefont {Kajiwara}, \citenamefont {Umezawa}, \citenamefont {Kawai},
  \citenamefont {Bauer}, \citenamefont {Maekawa},\ and\ \citenamefont
  {Saitoh}}]{Uchida2010}%
  \BibitemOpen
  \bibfield  {author} {\bibinfo {author} {\bibfnamefont {K.}~\bibnamefont
  {Uchida}}, \bibinfo {author} {\bibfnamefont {J.}~\bibnamefont {Xiao}},
  \bibinfo {author} {\bibfnamefont {H.}~\bibnamefont {Adachi}}, \bibinfo
  {author} {\bibfnamefont {J.}~\bibnamefont {Ohe}}, \bibinfo {author}
  {\bibfnamefont {S.}~\bibnamefont {Takahashi}}, \bibinfo {author}
  {\bibfnamefont {J.}~\bibnamefont {Ieda}}, \bibinfo {author} {\bibfnamefont
  {T.}~\bibnamefont {Ota}}, \bibinfo {author} {\bibfnamefont {Y.}~\bibnamefont
  {Kajiwara}}, \bibinfo {author} {\bibfnamefont {H.}~\bibnamefont {Umezawa}},
  \bibinfo {author} {\bibfnamefont {H.}~\bibnamefont {Kawai}}, \bibinfo
  {author} {\bibfnamefont {G.~E.~W.}\ \bibnamefont {Bauer}}, \bibinfo {author}
  {\bibfnamefont {S.}~\bibnamefont {Maekawa}}, \ and\ \bibinfo {author}
  {\bibfnamefont {E.}~\bibnamefont {Saitoh}},\ }\href {\doibase
  10.1038/nmat2856} {\bibfield  {journal} {\bibinfo  {journal} {Nature
  Materials}\ }\textbf {\bibinfo {volume} {9}},\ \bibinfo {pages} {894} (\bibinfo
  {year} {2010})}\BibitemShut {NoStop}%
\bibitem [{\citenamefont {Xiao}\ \emph {et~al.}(2010)\citenamefont {Xiao},
  \citenamefont {Bauer}, \citenamefont {Uchida}, \citenamefont {Saitoh},\ and\
  \citenamefont {Maekawa}}]{Xiao2010}%
  \BibitemOpen
  \bibfield  {author} {\bibinfo {author} {\bibfnamefont {J.}~\bibnamefont
  {Xiao}}, \bibinfo {author} {\bibfnamefont {G.~E.~W.}\ \bibnamefont {Bauer}},
  \bibinfo {author} {\bibfnamefont {K.~C.}~\bibnamefont {Uchida}}, \bibinfo
  {author} {\bibfnamefont {E.}~\bibnamefont {Saitoh}}, \ and\ \bibinfo {author}
  {\bibfnamefont {S.}~\bibnamefont {Maekawa}},\ }\href {\doibase
  10.1103/PhysRevB.81.214418} {\bibfield  {journal} {\bibinfo  {journal} {Phys.
  Rev. B}\ }\textbf {\bibinfo {volume} {81}},\ \bibinfo {pages} {214418}
  (\bibinfo {year} {2010})}\BibitemShut {NoStop}%
\bibitem [{\citenamefont {Adachi}\ \emph {et~al.}(2011)\citenamefont {Adachi},
  \citenamefont {Ohe}, \citenamefont {Takahashi},\ and\ \citenamefont
  {Maekawa}}]{Adachi2011}%
  \BibitemOpen
  \bibfield  {author} {\bibinfo {author} {\bibfnamefont {H.}~\bibnamefont
  {Adachi}}, \bibinfo {author} {\bibfnamefont {J.-i.}\ \bibnamefont {Ohe}},
  \bibinfo {author} {\bibfnamefont {S.}~\bibnamefont {Takahashi}}, \ and\
  \bibinfo {author} {\bibfnamefont {S.}~\bibnamefont {Maekawa}},\ }\href
  {\doibase 10.1103/PhysRevB.83.094410} {\bibfield  {journal} {\bibinfo
  {journal} {Phys. Rev. B}\ }\textbf {\bibinfo {volume} {83}},\ \bibinfo
  {pages} {094410} (\bibinfo {year} {2011})}\BibitemShut {NoStop}%
\bibitem [{\citenamefont {Kiselev}\ \emph {et~al.}(2003)\citenamefont
  {Kiselev}, \citenamefont {Sankey}, \citenamefont {Krivorotov}, \citenamefont
  {Emley}, \citenamefont {Schoelkopf}, \citenamefont {Buhrman},\ and\
  \citenamefont {Ralph}}]{Kiselev2003}%
  \BibitemOpen
  \bibfield  {author} {\bibinfo {author} {\bibfnamefont {S.~I.}\ \bibnamefont
  {Kiselev}}, \bibinfo {author} {\bibfnamefont {J.~C.}\ \bibnamefont {Sankey}},
  \bibinfo {author} {\bibfnamefont {I.~N.}\ \bibnamefont {Krivorotov}},
  \bibinfo {author} {\bibfnamefont {N.~C.}\ \bibnamefont {Emley}}, \bibinfo
  {author} {\bibfnamefont {R.~J.}\ \bibnamefont {Schoelkopf}}, \bibinfo
  {author} {\bibfnamefont {R.~A.}\ \bibnamefont {Buhrman}}, \ and\ \bibinfo
  {author} {\bibfnamefont {D.~C.}\ \bibnamefont {Ralph}},\ }\href {\doibase
  10.1038/nature01967} {\bibfield  {journal} {\bibinfo  {journal} {Nature}\
  }\textbf {\bibinfo {volume} {425}},\ \bibinfo {pages} {380} (\bibinfo {year}
  {2003})}\BibitemShut {NoStop}%
\bibitem [{\citenamefont {Houssameddine}\ \emph {et~al.}(2007)\citenamefont
  {Houssameddine}, \citenamefont {Ebels}, \citenamefont {Dela\"{e}t},
  \citenamefont {Rodmacq}, \citenamefont {Firastrau}, \citenamefont
  {Ponthenier}, \citenamefont {Brunet}, \citenamefont {Thirion}, \citenamefont
  {Michel}, \citenamefont {Prejbeanu-Buda}, \citenamefont {Cyrille},
  \citenamefont {Redon},\ and\ \citenamefont {Dieny}}]{Houssameddine2007}%
  \BibitemOpen
  \bibfield  {author} {\bibinfo {author} {\bibfnamefont {D.}~\bibnamefont
  {Houssameddine}}, \bibinfo {author} {\bibfnamefont {U.}~\bibnamefont
  {Ebels}}, \bibinfo {author} {\bibfnamefont {B.}~\bibnamefont {Dela\"{e}t}},
  \bibinfo {author} {\bibfnamefont {B.}~\bibnamefont {Rodmacq}}, \bibinfo
  {author} {\bibfnamefont {I.}~\bibnamefont {Firastrau}}, \bibinfo {author}
  {\bibfnamefont {F.}~\bibnamefont {Ponthenier}}, \bibinfo {author}
  {\bibfnamefont {M.}~\bibnamefont {Brunet}}, \bibinfo {author} {\bibfnamefont
  {C.}~\bibnamefont {Thirion}}, \bibinfo {author} {\bibfnamefont {J.-P.}\
  \bibnamefont {Michel}}, \bibinfo {author} {\bibfnamefont {L.}~\bibnamefont
  {Prejbeanu-Buda}}, \bibinfo {author} {\bibfnamefont {M.-C.}\ \bibnamefont
  {Cyrille}}, \bibinfo {author} {\bibfnamefont {O.}~\bibnamefont {Redon}}, \
  and\ \bibinfo {author} {\bibfnamefont {B.}~\bibnamefont {Dieny}},\ }\href
  {\doibase 10.1038/nmat1905} {\bibfield  {journal} {\bibinfo  {journal}
  {Nature materials}\ }\textbf {\bibinfo {volume} {6}},\ \bibinfo {pages} {441}
  (\bibinfo {year} {2007})}\BibitemShut {NoStop}%
\bibitem [{\citenamefont {Brataas}\ \emph {et~al.}(2002)\citenamefont
  {Brataas}, \citenamefont {Tserkovnyak}, \citenamefont {Bauer},\ and\
  \citenamefont {Halperin}}]{Brataas2002}%
  \BibitemOpen
  \bibfield  {author} {\bibinfo {author} {\bibfnamefont {A.}~\bibnamefont
  {Brataas}}, \bibinfo {author} {\bibfnamefont {Y.}~\bibnamefont
  {Tserkovnyak}}, \bibinfo {author} {\bibfnamefont {G.~E.~W.}~\bibnamefont {Bauer}},
  \ and\ \bibinfo {author} {\bibfnamefont {B.~I.}~\bibnamefont {Halperin}},\
  }\href {\doibase 10.1103/PhysRevB.66.060404} {\bibfield  {journal} {\bibinfo
  {journal} {Physical Review B}\ }\textbf {\bibinfo {volume} {66}},\ \bibinfo
  {pages} {060404} (\bibinfo {year} {2002})}\BibitemShut {NoStop}%
\bibitem [{\citenamefont {Tserkovnyak}\ \emph
  {et~al.}(2002{\natexlab{a}})\citenamefont {Tserkovnyak}, \citenamefont
  {Brataas},\ and\ \citenamefont {Bauer}}]{Tserkovnyak2002a}%
  \BibitemOpen
  \bibfield  {author} {\bibinfo {author} {\bibfnamefont {Y.}~\bibnamefont
  {Tserkovnyak}}, \bibinfo {author} {\bibfnamefont {A.}~\bibnamefont
  {Brataas}}, \ and\ \bibinfo {author} {\bibfnamefont {G.~E.~W.}\ \bibnamefont
  {Bauer}},\ }\href {\doibase 10.1103/PhysRevB.66.224403} {\bibfield  {journal}
  {\bibinfo  {journal} {Phys. Rev. B}\ }\textbf {\bibinfo {volume} {66}},\
  \bibinfo {pages} {224403} (\bibinfo {year} {2002}{\natexlab{a}})}\BibitemShut
  {NoStop}%
\bibitem [{\citenamefont {Tserkovnyak}\ \emph
  {et~al.}(2002{\natexlab{b}})\citenamefont {Tserkovnyak}, \citenamefont
  {Brataas},\ and\ \citenamefont {Bauer}}]{Tserkovnyak2002}%
  \BibitemOpen
  \bibfield  {author} {\bibinfo {author} {\bibfnamefont {Y.}~\bibnamefont
  {Tserkovnyak}}, \bibinfo {author} {\bibfnamefont {A.}~\bibnamefont
  {Brataas}}, \ and\ \bibinfo {author} {\bibfnamefont {G.~E.~W.}\ \bibnamefont
  {Bauer}},\ }\href {\doibase 10.1103/PhysRevLett.88.117601} {\bibfield
  {journal} {\bibinfo  {journal} {Phys. Rev. Lett.}\ }\textbf {\bibinfo
  {volume} {88}},\ \bibinfo {pages} {117601} (\bibinfo {year}
  {2002}{\natexlab{b}})}\BibitemShut {NoStop}%
\bibitem [{\citenamefont {Brataas}\ \emph {et~al.}(2008)\citenamefont
  {Brataas}, \citenamefont {Tserkovnyak},\ and\ \citenamefont
  {Bauer}}]{Brataas2008}%
  \BibitemOpen
  \bibfield  {author} {\bibinfo {author} {\bibfnamefont {A.}~\bibnamefont
  {Brataas}}, \bibinfo {author} {\bibfnamefont {Y.}~\bibnamefont
  {Tserkovnyak}}, \ and\ \bibinfo {author} {\bibfnamefont {G.~E.~W.}\
  \bibnamefont {Bauer}},\ }\href {\doibase 10.1103/PhysRevLett.101.037207}
  {\bibfield  {journal} {\bibinfo  {journal} {Phys. Rev. Lett.}\ }\textbf
  {\bibinfo {volume} {101}},\ \bibinfo {pages} {037207} (\bibinfo {year}
  {2008})}\BibitemShut {NoStop}%
\bibitem [{\citenamefont {Meckenstock}(2008)}]{Meckenstock2008}%
  \BibitemOpen
  \bibfield  {author} {\bibinfo {author} {\bibfnamefont {R.}~\bibnamefont
  {Meckenstock}},\ }\href {\doibase DOI:10.1063/1.2908445} {\bibfield
  {journal} {\bibinfo  {journal} {Rev. Sci. Instrum.}\ }\textbf {\bibinfo
  {volume} {79}},\ \bibinfo {pages} {041101} (\bibinfo {year}
  {2008})}\BibitemShut {NoStop}%
\bibitem [{\citenamefont {Sakran}\ \emph {et~al.}(2004)\citenamefont {Sakran},
  \citenamefont {Copty}, \citenamefont {Golosovsky}, \citenamefont {Davidov},\
  and\ \citenamefont {Monod}}]{Sakran2004}%
  \BibitemOpen
  \bibfield  {author} {\bibinfo {author} {\bibfnamefont {F.}~\bibnamefont
  {Sakran}}, \bibinfo {author} {\bibfnamefont {A.}~\bibnamefont {Copty}},
  \bibinfo {author} {\bibfnamefont {M.}~\bibnamefont {Golosovsky}}, \bibinfo
  {author} {\bibfnamefont {D.}~\bibnamefont {Davidov}}, \ and\ \bibinfo
  {author} {\bibfnamefont {P.}~\bibnamefont {Monod}},\ }\href {\doibase
  DOI:10.1063/1.1756682} {\bibfield  {journal} {\bibinfo  {journal} {Applied
  Physics Letters}\ }\textbf {\bibinfo {volume} {84}},\ \bibinfo {pages} {4499}
  (\bibinfo {year} {2004})}\BibitemShut {NoStop}%
\bibitem [{\citenamefont {Coey}(2010)}]{coey2010magnetism}%
  \BibitemOpen
  \bibfield  {author} {\bibinfo {author} {\bibfnamefont {J.}~\bibnamefont
  {Coey}},\ }\href {http://books.google.com/books?id=\_olKSAAACAAJ} {\emph
  {\bibinfo {title} {Magnetism and magnetic materials}}}\ (\bibinfo
  {publisher} {Cambridge University Press},\ \bibinfo {year}
  {2010})\BibitemShut {NoStop}%
\bibitem [{\citenamefont {Costache}\ \emph
  {et~al.}(2006{\natexlab{a}})\citenamefont {Costache}, \citenamefont {Watts},
  \citenamefont {Sladkov}, \citenamefont {van~der Wal},\ and\ \citenamefont
  {van Wees}}]{Costache2006a}%
  \BibitemOpen
  \bibfield  {author} {\bibinfo {author} {\bibfnamefont {M.~V.}\ \bibnamefont
  {Costache}}, \bibinfo {author} {\bibfnamefont {S.~M.}\ \bibnamefont {Watts}},
  \bibinfo {author} {\bibfnamefont {M.}~\bibnamefont {Sladkov}}, \bibinfo
  {author} {\bibfnamefont {C.~H.}\ \bibnamefont {van~der Wal}}, \ and\ \bibinfo
  {author} {\bibfnamefont {B.~J.}\ \bibnamefont {van Wees}},\ }\href {\doibase
  10.1063/1.2400058} {\bibfield  {journal} {\bibinfo  {journal} {Applied
  Physics Letters}\ }\textbf {\bibinfo {volume} {89}},\ \bibinfo {pages}
  {232115} (\bibinfo {year} {2006}{\natexlab{a}})}\BibitemShut {NoStop}%
\bibitem [{\citenamefont {Bakker}\ \emph {et~al.}(2010)\citenamefont {Bakker},
  \citenamefont {Slachter}, \citenamefont {Adam},\ and\ \citenamefont {van
  Wees}}]{Bakker2010}%
  \BibitemOpen
  \bibfield  {author} {\bibinfo {author} {\bibfnamefont {F.~L.}\ \bibnamefont
  {Bakker}}, \bibinfo {author} {\bibfnamefont {A.}~\bibnamefont {Slachter}},
  \bibinfo {author} {\bibfnamefont {J.-P.}\ \bibnamefont {Adam}}, \ and\
  \bibinfo {author} {\bibfnamefont {B.~J.}\ \bibnamefont {van Wees}},\ }\href
  {\doibase 10.1103/PhysRevLett.105.136601} {\bibfield  {journal} {\bibinfo
  {journal} {Physical Review Letters}\ }\textbf {\bibinfo {volume} {105}},\
  \bibinfo {pages} {136601} (\bibinfo {year} {2010})}\BibitemShut {NoStop}%
\bibitem [{\citenamefont {{Slachter}}\ \emph {et~al.}(2011)\citenamefont
  {{Slachter}}, \citenamefont {{Bakker}},\ and\ \citenamefont {{van
  Wees}}}]{Slachter2011}%
  \BibitemOpen
  \bibfield  {author} {\bibinfo {author} {\bibfnamefont {A.}~\bibnamefont
  {{Slachter}}}, \bibinfo {author} {\bibfnamefont {F.~L.}\ \bibnamefont
  {{Bakker}}}, \ and\ \bibinfo {author} {\bibfnamefont {B.~J.}\ \bibnamefont
  {{van Wees}}},\ }\href@noop {} {\bibfield  {journal} {\bibinfo  {journal}
  {ArXiv e-prints}\ } (\bibinfo {year} {2011})},\ \Eprint
  {http://arxiv.org/abs/1107.3290} {arXiv:1107.3290 [cond-mat.mes-hall]}
  \BibitemShut {NoStop}%
\bibitem [{\citenamefont {Kittel}(1995)}]{Kittel1995}%
  \BibitemOpen
  \bibfield  {author} {\bibinfo {author} {\bibfnamefont {C.}~\bibnamefont
  {Kittel}},\ }\href@noop {} {\emph {\bibinfo {title} {Introduction to Solid
  State Physics -7th ed.}}}\ (\bibinfo  {publisher} {John Wiley \& Sons, Inc.,
  New York},\ \bibinfo {year} {1995})\BibitemShut {NoStop}%
\bibitem [{\citenamefont {Wang}\ \emph {et~al.}(2006)\citenamefont {Wang},
  \citenamefont {Bauer}, \citenamefont {van Wees}, \citenamefont {Brataas},\
  and\ \citenamefont {Tserkovnyak}}]{Wang2006}%
  \BibitemOpen
  \bibfield  {author} {\bibinfo {author} {\bibfnamefont {X.}~\bibnamefont
  {Wang}}, \bibinfo {author} {\bibfnamefont {G.~E.~W.}\ \bibnamefont {Bauer}},
  \bibinfo {author} {\bibfnamefont {B.~J.}\ \bibnamefont {van Wees}}, \bibinfo
  {author} {\bibfnamefont {A.}~\bibnamefont {Brataas}}, \ and\ \bibinfo
  {author} {\bibfnamefont {Y.}~\bibnamefont {Tserkovnyak}},\ }\href {\doibase
  10.1103/PhysRevLett.97.216602} {\bibfield  {journal} {\bibinfo  {journal}
  {Phys. Rev. Lett.}\ }\textbf {\bibinfo {volume} {97}},\ \bibinfo {pages}
  {216602} (\bibinfo {year} {2006})}\BibitemShut {NoStop}%
\bibitem [{\citenamefont {Costache}\ \emph
  {et~al.}(2006{\natexlab{b}})\citenamefont {Costache}, \citenamefont
  {Sladkov}, \citenamefont {Watts}, \citenamefont {van~der Wal},\ and\
  \citenamefont {van Wees}}]{Costache2006b}%
  \BibitemOpen
  \bibfield  {author} {\bibinfo {author} {\bibfnamefont {M.~V.}\ \bibnamefont
  {Costache}}, \bibinfo {author} {\bibfnamefont {M.}~\bibnamefont {Sladkov}},
  \bibinfo {author} {\bibfnamefont {S.~M.}~\bibnamefont {Watts}}, \bibinfo
  {author} {\bibfnamefont {C.~H.}\ \bibnamefont {van~der Wal}}, \ and\ \bibinfo
  {author} {\bibfnamefont {B.~J.}\ \bibnamefont {van Wees}},\ }\href {\doibase
  10.1103/PhysRevLett.97.216603} {\bibfield  {journal} {\bibinfo  {journal}
  {Physical Review Letters}\ }\textbf {\bibinfo {volume} {97}},\ \bibinfo
  {pages} {216603} (\bibinfo {year} {2006}{\natexlab{b}})}\BibitemShut {NoStop}%
\end{thebibliography}
\end{document}